\documentclass[epsfig,graphics,twocolumn,floatfix,mathbbm,prl]{revtex4}
\usepackage{graphicx}

\newcommand{\PrYSO}{Pr$^{3+}$:Y$_2$SiO$_5$}
\newcommand{\Prtrans}{$^3$H$_4\rightarrow^1$D$_2$}
\newcommand{\abstrans}{$|g\rangle\leftrightarrow|e\rangle$}
\newcommand{\spintrans}{$|s\rangle\leftrightarrow|e\rangle$}

\begin{document}

\title{Demonstration of atomic frequency comb memory for light with spin-wave storage}

\author{Mikael Afzelius$^1$}
\email{mikael.afzelius@unige.ch}
\author{Imam Usmani$^1$}
\author{Atia Amari$^2$}
\author{Bj\"{o}rn Lauritzen$^1$}
\author{Andreas Walther$^2$}
\author{Christoph Simon$^1$}
\author{Nicolas Sangouard$^1$}
\author{Ji\v{r}\'i Min\'{a}\v{r}$^1$}
\author{Hugues de Riedmatten$^1$}
\author{Nicolas Gisin$^1$}
\author{Stefan Kr\"{o}ll$^2$}

\address{$^1$Group of Applied Physics, University of Geneva, CH-1211 Geneva 4, Switzerland}%
\address{$^2$Department of Physics, Lund University, SE-22100 Lund, Sweden}%

\date{\today}

\begin{abstract}
We present a light-storage experiment in a praseodymium-doped crystal where the light is mapped onto an inhomogeneously broadened optical transition shaped into an atomic frequency comb. After absorption of the light the optical excitation is converted into a spin-wave excitation by a control pulse. A second control pulse reads the memory (on-demand) by reconverting the spin-wave excitation to an optical one, where the comb structure causes a photon-echo type rephasing of the dipole moments and directional retrieval of the light. This combination of photon echo and spin-wave storage allows us to store sub-microsecond (450ns) pulses for up to 20 $\mu$s. The scheme has a high potential for storing multiple temporal modes in the single photon regime, which is an important resource for future long-distance quantum communication based on quantum repeaters.
\end{abstract}

\maketitle

A quantum memory (QM) for photons is a light-matter interface that can achieve a coherent and reversible transfer of quantum information between a light field and a material system \cite{Hammerer2008}. A QM should enable efficient, high-fidelity storage of non-classical states of light, which is a key resource for future quantum networks, particularly in quantum repeaters \cite{Briegel1998,Duan2001,Simon2007,Collins2007,Sangouard2009} that have the potential for distributing entangled states over long distances for quantum communication tasks. In order to achieve reasonable entanglement distribution rates, it has been shown that some type of multiplexing is required \cite{Simon2007,Collins2007}, using for instance independent frequency, spatial or temporal modes (multimode QM).

Several types of light-matter interactions have been proposed for building a QM, for instance electromagnetically induced transparency \cite{Fleischhauer2000,Chaneliere2005,Eisaman2005,Choi2008}, Raman interactions \cite{Kozhekin2000,Julsgaard2004,Nunn2007,Hetet2009}, or photon-echo techniques \cite{Moiseev2001,Kraus2006,Alexander2006,Hetet2008b,Tittel2008,Afzelius2009,deRiedmatten2008,Chaneliere2009}. Photon echo techniques in rare-earth-ion doped crystals have an especially high multimode capacity for storing classical light \cite{Lin1995}. Classical photon echoes are not useful, however, for single-photon storage due to inherent noise problems due to unwanted spontaneous and stimulated emission processes when storing light on a single photon level \cite{Ruggiero2009}. The photon-echo QM based on controlled reversible inhomogeneous broadening \cite{Moiseev2001,Kraus2006,Alexander2006,Hetet2008b,Tittel2008} is free of these noise problems. But this technique has a lower time-multiplexing capacity than classical photon echoes, for a given optical depth, due to loss of storage efficiency as the controlled frequency bandwidth is increased \cite{Afzelius2009,Nunn2008a}. Some of us recently proposed a photon-echo type QM based on an atomic frequency comb (AFC) \cite{Afzelius2009} that has a storage efficiency independent of the bandwidth, allowing optimal use of the inhomogeneous broadening of rare-earth-doped crystals. An AFC memory has the potential for providing multimode storage capacity \cite{Afzelius2009,Nunn2008a} crucial to quantum repeaters. In a first experiment \cite{deRiedmatten2008} based on this scheme we performed a light-matter interface at the single-photon level. However, the light was retrieved after a predetermined storage time, while for quantum repeaters it is crucial to be able to choose the time of the memory readout (on-demand readout). Here we present the first light-storage experiment where an AFC is used in combination with reversible transfer of the excitation to a spin state \cite{Afzelius2009}, resulting in on-demand readout and storage times longer than 20$\mu$s.

\begin{figure}
    \centering
    \includegraphics[width=.45\textwidth]{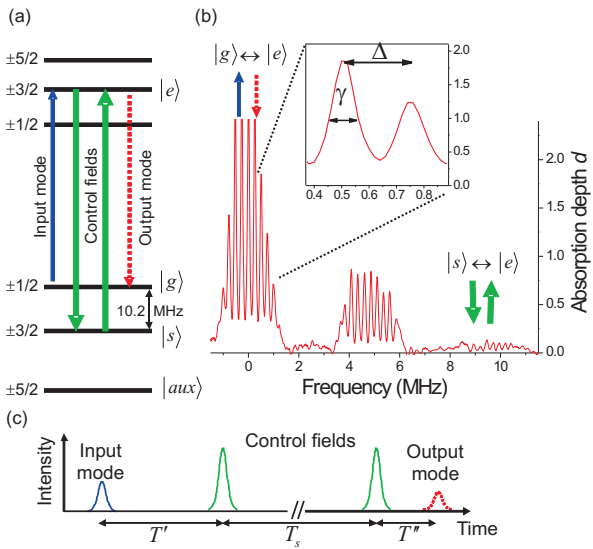}
    \caption{(Color online) (a) The experiment was performed on the \Prtrans\ transition in Pr$^{3+}$. The ground and excited state manifolds both have three hyperfine levels denoted M$_I$=$\pm$1/2,$\pm$3/2,$\pm$5/2. The three-level lambda system was formed by the levels labeled $|g\rangle$, $|s\rangle$ and $|e\rangle$, following the notation in \cite{Afzelius2009}. (b) Experimental absorption spectrum showing the AFC on the \abstrans\ transition created within a 18 MHz wide transmission hole using the spectral holeburning sequence described in the text. Here the comb consists of 9 peaks with spacing $\Delta$=250 kHz. The holeburning sequence also empties the $|s\rangle$ level, whereas $|aux\rangle$ is used for population storage. Note that the second AFC in the centre of the spectrum is due to the weaker transition from the ground state $\pm$1/2 to the excited state $\pm$5/2. (c) The pulse sequence showing the input pulse to store, the two control fields for the back-and-forth transfer to the spin state, and the retrieved output pulse.}
    \label{fig_spectrum}
\end{figure}

We start by giving a brief description of the AFC QM. The underlying idea is to shape an inhomogeneously broadened optical transition $|g\rangle\rightarrow|e\rangle$ into a periodic series of narrow and highly absorbing peaks with periodicity $\Delta$, see Fig. \ref{fig_spectrum}. A photon with a bandwidth that is matched to the width of the AFC structure is then stored as an optical excitation delocalized over the peaks, which we can write as $|\psi\rangle=\sum_{j=1}^N c_j e^{i\delta_j
t}e^{-ikz_j} |g_1\cdot\cdot\cdot e_j\cdot\cdot\cdot g_N \rangle$ where $N$ is the number of atoms in the AFC, $\delta_j$ is the detuning of atom $j$ with respect to the laser frequency, $z_j$ is the position, $k$ is the wave-number of the light field, and
the amplitudes $c_j$ depend on the frequency and on the spatial position of the particular atom $j$. The terms in this large superposition state accumulate different phases due to the inhomogeneous distribution of atomic resonance frequencies, resulting in a loss of the initially strong collective coupling to the light mode. But the periodic AFC peak separation $\Delta$ leads to a rephasing of the terms after a time $1/\Delta$, which restores the strong collective coupling, leading to a photon-echo type re-emission \cite{pecho}, the AFC echo. The narrow and highly absorbing peaks can theoretically absorb all the light and completely emit the energy in the AFC echo \cite{Afzelius2009}. A large number of peaks leads to a high multimode capacity \cite{Afzelius2009,Nunn2008a}, since the number of independent temporal modes that can be stored is given by the ratio of the AFC bandwidth to the peak separation (assuming the photon bandwidth is less than the AFC bandwidth).

The few reported AFC experiments \cite{deRiedmatten2008,Chaneliere2009} have been investigating the physics of the optical AFC echo, where the memory storage time is predetermined by the periodicity.  In the original proposal \cite{Afzelius2009} on-demand readout and longer storage times rely on reversible transfer of the optical excitation to a long-lived spin state by strong control pulses (see Fig. \ref{fig_spectrum}). If we imagine a perfect $\pi$-pulse applied at time $T'$, each term in the superposition state becomes $|g_1\cdot\cdot\cdot s_j\cdot\cdot\cdot g_N \rangle$, thus we have a single collective spin-wave excitation. If we assume that the spin transition is homogeneously broadened, then each term is frozen with the phase term $e^{i\delta_j T'}$ due to the time spent in the excited state \cite{Afzelius2009}. In practice, inhomogeneous spin broadening adds other phase factors and reduces the collective spin wave, as will be discussed below. After a time $T_s$ in the spin state, another control pulse transfers the excitation back to the excited state and the AFC evolution resumes, leading to a re-emission after a total storage time $T'+T_s+T''$ where $T'+T''=1/\Delta$.

\begin{figure}[t]
    \centering
    \includegraphics[width=.45\textwidth]{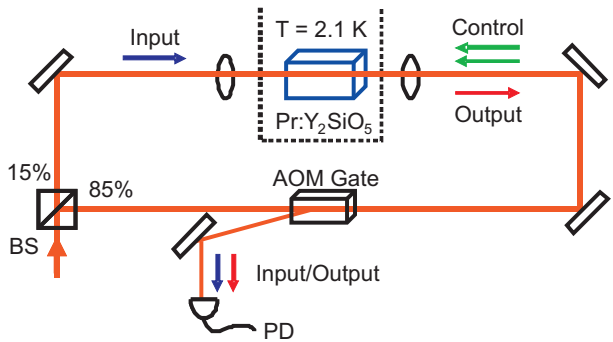}
    \caption{(Color online) The experimental setup. The spectral hole burning and storage pulse sequences were created using a frequency-stabilized laser and acousto-optic modulators similar to the setup in \cite{Rippe2005} (not shown). A beamsplitter (BS) split the light into a strong and a weak beam, whose two modes were overlapped in the crystal cooled to 2.1 K. The detected output pulse propagating to the right originated from the combination of a weak input pulse incident on the crystal from the left, and two strong control pulses incident from the right. Note that the signal disappeared when the control pulses were blocked directly after the BS. An AOM was used to direct only the transmitted input pulse and the output pulse onto the photodiode (PD), effectively working as a detector gate.}
    \label{fig_setup}
\end{figure}

 The storage material used in this experiment is a praseodymium-doped Y$_2$SiO$_5$ crystal (Pr$^{3+}$ concentration of 0.05\%) with an optical transition at 606 nm. The optical homogeneous linewidth at cryogenic temperature is around 1 kHz, whereas the inhomogeneous broadening is about 5 GHz \cite{Equall1995}. The ground and excited states both have a hyperfine manifold consisting of three closely spaced levels [Fig. \ref{fig_spectrum}(a)], assuming no applied magnetic field. Three ground-state levels are necessary for the experiment, see below, which was our main motivation for choosing Pr$^{3+}$. The different hyperfine transitions are usually hidden within the large inhomogeneous broadening. By spectral hole burning techniques one can, however, isolate a sub-ensemble of atoms whose different hyperfine transitions can be unambiguously excited. This distillation technique, which we will summarize here, has been the subject of several papers \cite{Nilsson2004,Rippe2005,Rippe2008}. A laser beam whose frequency is swept pumps atoms from ground levels $|g\rangle$ and $|s\rangle$ to the auxiliary storage level $|aux\rangle$, see Fig. \ref{fig_spectrum}, which creates a wide transmission window within the inhomogeneous profile. In the next step a narrow absorption peak is created in the hole by coherently transferring back atoms, within a narrow frequency range, from $|aux\rangle$ to $|g\rangle$ \cite{Rippe2005}. In this experiment we extended this method by transferring back atoms at different frequencies to create a frequency comb. In Fig. \ref{fig_spectrum}(b) we show the absorption spectrum recorded after the preparation sequence. The AFC created on the \abstrans\ transition is clearly visible. The $|s\rangle$ level is used for the spin-wave storage, which means that the control pulses will be applied on the \spintrans\ transition displaced by 10.2 MHz with respect to the \abstrans\ transition. Note that the absorption spectrum in Fig. \ref{fig_spectrum}(b) is shown only for visualization purposes, since the fast frequency scan method \cite{Chang2005} we used leads to distortions for absorption depths $d\gtrsim$2, thus preventing us from accurately measuring the comb parameters. The experimental setup is shown in Fig. \ref{fig_setup}. The control pulses were both counterpropagating with respect to the input pulse. By phase matching condition the output pulse then copropagates with the input signal \cite{Afzelius2009}. Using this configuration we could reduce noise due to off-resonant free-induction decay emission produced by the strong control pulses.

\begin{figure}[t]
    \centering
    \includegraphics[width=.45\textwidth]{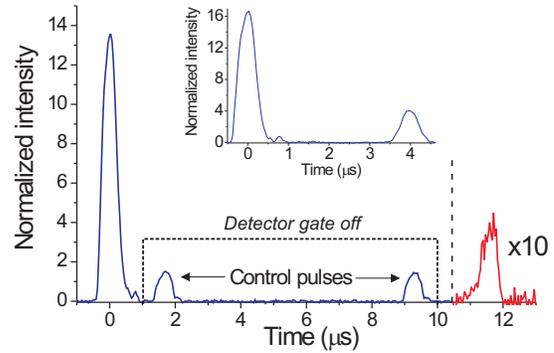}
    \caption{(Color online) Storage in the spin state using two control pulses. Shown are (from left) the partially transmitted input pulse, control pulses (strongly attenuated by the closed optical gate), and the output pulse (magnified by a factor 10 for clarity). Here $1/\Delta$=4 $\mu$s and $T_s$ = 7.6 $\mu$s, resulting in a total storage time $1/\Delta+T_s$=11.6 $\mu$s. Inset: The AFC echo observed at $1/\Delta$=4 $\mu$s when the control pulses are not applied. Note that the vertical scales have been normalized to 100 with respect to the input pulse before the crystal, thus these yield (in percent) the efficiency for the echo and transmission coefficient for the input pulse.}
    \label{fig_w_wo_control}
\end{figure}

In a preliminary experiment we investigate the AFC echo on the \abstrans\ transition without applying the control pulses, see inset in Fig. \ref{fig_w_wo_control}. This allows us to optimize the relevant comb parameters in order to obtain a strong echo. The input pulse duration was set to 450ns so that the bandwidth is entirely contained in the 2 MHz wide AFC. The efficiency of this AFC echo, which we define as the ratio of the AFC echo area to the input pulse area, depend on the shape of the AFC \cite{Afzelius2009}. Two critical parameters are the peak absorption depth $d$ and the finesse defined as $F=\Delta/\gamma$ where $\gamma$ is the full-width at half maximum of a peak. For instance, a high finesse leads to low decoherence during the storage time $1/\Delta$, but also to a lower effective absorption $\sim d/F$ of the input pulse \cite{Afzelius2009}. The peak absorption $d$ could be controlled by the power of the laser beam creating the peaks in the transmission hole, but a high power also had an impact on the finesse by causing powerbroadening of the peaks. The peak width was also limited by laser frequency stability, resulting in typical widths of $\gamma\approx$100 kHz. For a periodicity of $\Delta$=1 MHz, the optimized efficiency was about 15$\%$. The delay of 1 $\mu$s was not sufficiently long, however, for applying the control pulses before the emission of the AFC echo. We therefore set the periodicity to $\Delta$=250 kHz giving us 4 $\mu$s to apply the control pulses. The closer spacing of the peaks lowered the finesse of the comb, thus lowering the efficiency to $\sim$5\% (see inset in Fig. \ref{fig_w_wo_control}). This efficiency is in reasonable agreement with numerical simulations using the experimentally estimated peak absorption depth and finesse of the comb. A more detailed analysis would require a more accurate measurement of the comb structure.

In Fig. \ref{fig_w_wo_control} we show the main result where two control pulses are applied on \spintrans\ to transfer the excitation to the $|s\rangle$ hyperfine level. The retrieved pulse is clearly observed above the noise level. This realizes a true storage of the input pulses, with on-demand readout. Thus, the control pulses provide a mechanism for momentarily interrupting the predetermined AFC evolution \cite{Afzelius2009}. We tested this mechanism in detail by varying the time at which the first control pulse was applied $T'$=(1.17, 1.63, 2.23)$\mu$s (cf. Fig. \ref{fig_spectrum}(c) for notation). This resulted in different measured durations $T''$=(2.84, 2.41, 1.85)$\mu$s. The total time spent in the excited state $|e\rangle$, however, is constant (within the measurement error), $T'+T''$=(4.01, 4.04, 4.08)$\mu$s, corresponding to the expected $1/\Delta$.

\begin{figure}[b]
    \centering
    \includegraphics[width=.50\textwidth]{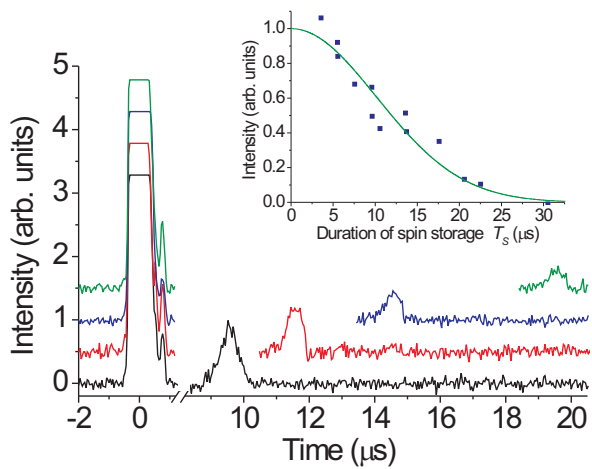}
    \caption{(Color online) Experimental traces for spin storage times $T_s$ = 5.6, 7.6, 10.6, and 15.6 $\mu$s. All other parameters are the same as those in Fig. \ref{fig_w_wo_control}. The input pulses are superimposed to the left (truncated) and the output pulses for different $T_s$ are seen to the right. The leakage of the control pulses through the optical gate is not shown. For clarity there is also a break in the horizontal scale. The decay of the signal (see inset) as a function of $T_s$ is due to inhomogeneous spin dephasing. The solid curve is a fitted Gaussian function corresponding to 26 kHz (full-width at half maximum) spin broadening.}
    \label{fig_multi_delays_decay}
\end{figure}

In Fig. \ref{fig_multi_delays_decay} we show storage experiments where the spin-wave storage time $T_s$ is varied. The output signal is clearly visible up to 20$\mu$s of total storage time ($1/\Delta+T_s$). The exponential decay of the output signal as a function of $T_s$ can be attributed to inhomogeneous spin dephasing, corresponding to an inhomogeneous broadening of 26 kHz consistent with previous measurements \cite{Ham2003}. We point out that this can be compensated for by spin echo techniques. With such techniques Longdell et. al \cite{Longdell2005} stopped light during $>$1 second in a \PrYSO\ crystal using electromagnetically induced transparency.

We now discuss the total storage efficiency $\eta$. Clearly $\eta$ is bounded by the AFC echo efficiency $\eta_e$=4-5$\%$ for $1/\Delta$=4$\mu$s (cf. Fig. \ref{fig_w_wo_control} inset). This can be improved by increasing the finesse and optical depth (see discussion above). The spin-wave storage further decreases the efficiency. By extrapolating the experimentally measured $\eta$ to the limit $T_s\rightarrow0$ we find that $\eta$=0.5-1$\%$, which is a value independent of spin dephasing. The effect of the control pulses can now be understood by a simple model. We assume that one pulse has a single-atom $|s\rangle \rightarrow |e\rangle$ transfer efficiency $\eta_T$, which is constant as a function of detuning. Then the application of the two control pulses reduce the efficiency to $\eta=\eta_e\cdot\eta_T^2$. Numerical simulations using 3-level Maxwell-Bloch equations show that this simple model is correct if the control pulses do not induce any additional phase decoherence. From the experimental values given above we thus find $\eta_T\approx0.30-0.45$.

In order to better understand this value we have performed numerical calculations. The control pulses were two identical complex hyperbolic secant pulses, which can achieve efficient, broadband transfer of population \cite{Rippe2005} and coherence \cite{Rippe2008}. Our pulses had duration 600 ns and Rabi frequency $\sim$1.2 MHz (close to the maximal value in this experiment), the frequency chirp being 2 MHz. These values were found by empirically optimizing the size of the output pulse. Using the numerical model we find $\eta_T$=0.75, also averaged over the bandwidth of the AFC. Based on this we would expect a total efficiency $\eta=5\%\cdot0.75^2=2.8\%$ using the simplified model above, which is significantly higher than what we observe. The most probable explanation for the discrepancy is imperfect spatial mode overlap between the counterpropagating input and control pulses (see Fig. \ref{fig_setup}). A larger control beam would make a more spatially uniform Rabi frequency and would facilitate mode alignment. We also note that the theoretical $\eta_T$ can be improved by increasing the Rabi frequency (by a factor of two) and adapting the duration and chirp of the pulse in order to achieve an efficient ($\eta_T\geq95\%$) transfer.

\begin{figure}
    \centering
    \includegraphics[width=.45\textwidth]{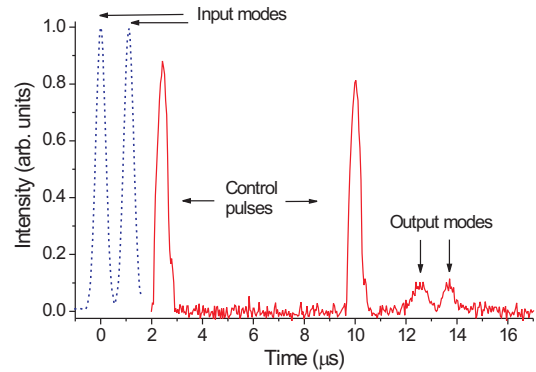}
    \caption{(Color online) Storage of two temporal input modes. The normalized input modes shown here (dashed line) are recorded before the crystal using a reference detector. The AFC was prepared with a periodicity of $\Delta$=200 kHz, corresponding to 5 $\mu$s storage time on the optical transition. The spin storage time was set to $T_s$ = 7.6 $\mu$s, resulting in a total storage time of $1/\Delta+T_s$=12.6 $\mu$s.}
    \label{fig_2_modes}
\end{figure}

We finally show an example of storage of two temporal modes, see Fig. \ref{fig_2_modes}. Note that both modes are stored with the same efficiency, which is a particular feature of the AFC memory due to the fact that each mode spend the same total time ($1/\Delta+T_s$) in the memory. The number of modes we could store was mainly limited by the number of peaks $N_p$ that could be created in the AFC, since the number of input modes one can store is proportional to $N_p$ \cite{Afzelius2009}. $N_p$ could be increased by making narrower peaks and/or a wider AFC. The width is currently limited by the separation of the hyperfine transitions (cf. Fig. \ref{fig_spectrum}), which could be increased via the nuclear Zeeman effect. The most significant improvement can be made by creating narrower peaks, which in principle can approach the homogeneous linewidth of about 1 kHz. A multimode storage capacity in the range of tens of modes appears feasible. Other rare-earth-ion-doped crystals have even higher multi-mode potential, for instance Eu$^{3+}$:Y$_2$SiO$_5$ as discussed in Ref. \cite{Afzelius2009}.

In conclusion, we have demonstrated the first light-storage experiment combining a photon-echo technique based on an atomic frequency comb and spin-wave storage. Using this method we stored optical sub-microsecond (450 ns) pulses for up to 20 $\mu$s as a spin-wave in \PrYSO. This optical bandwidth is more than one order of magnitude higher than previous stopped-light experiments demonstrated in rare-earth crystals \cite{Longdell2005,Turukhin2002}. The spin-storage time could be greatly extended by spin echo techniques \cite{Longdell2005}. Our method can be used for storing light at the single photon level \cite{deRiedmatten2008}, where it could provide high temporal multimode storage for quantum repeaters.

\indent The authors acknowledge useful discussions with Pavel Sekatski. The work
was supported by the Swiss NCCR Quantum Photonics, the Swedish Research Council, the Knut and Alice Wallenberg Foundation, the ERC Advanced Grant QORE, the Lund Laser Center, and the EC projects Qubit
Applications (QAP) and FP7 grant n$^\circ$ 228334.

\bibliographystyle{unsrt}

\end{document}